\documentclass[12pt]{article}

\usepackage{mathptmx}

\usepackage{amsmath,amssymb}

\newtheorem{lemma}{Lemma}[section]

\newtheorem{theorem}{Theorem}[section]
\newtheorem{proposition}{Proposition}[section]

\newcommand{\proof}{ $\triangleright$\quad}
\newcommand{\qed}{\hfill $\triangleleft$}

\begin{document}

\title {Energy-mass spectrum of Yang-Mills bosons is infinite and discrete}

\author{Alexander  Dynin\\
\textit{\small Department of Mathematics, Ohio State University}\\
\textit{\small Columbus, OH 43210, USA}, \texttt{\small dynin@math.ohio-state.edu}}

\maketitle

\begin{abstract}
A non-perturbative anti-normal quantization  of  relativistic Yang-Mills fields with a compact  semisimple gauge  group entails  an infinite discrete bosonic energy-mass  spectrum of  gauge bosons in the framework of Gelfand nuclear triples. The quantum   spectrum is bounded from below  and has a positive   mass gap. The spectrum is both Poincare and gauge invariant.\footnote
{2010 \emph{MSC}: Primary 81T08, 81T13; Secondary 60H40, 46G20.

}

\end{abstract}

\begin{center}
\emph{In memoriam\\  F. A. Berezin (1931-1980), I. M. Gelfand (1913 - 2009),\\ and I. E Segal  (1918 - 1998)}
\end{center}

\section{Introduction}
\subsection{Yang-Mills  problem}

\smallskip

This  paper  offers a mathematically rigorous  quantum Yang-Mills theory on Minkowski 4-space with an   infinite and discrete  energy-mass quantum bosonic spectrum  for any compact  semisimple gauge  group. This Lagrangian  theory is non-perturbative and ghostless. It is also Higgsless but requires an infinite anti-normal renormalization.

\smallskip
As an application, the theory gives  \emph{a} solution for the  7th  of the Clay Mathematics Institute  "Millennium Prize Problems" (\textsc{jaffe-witten} \cite{Clay-00}):
\begin{quotation}
\textsl{Prove that for any compact (semi-)simple global gauge group, a nontrivial  quantum Yang-Mills theory exists  on $\mathbb{R}^{1+3}$ and has a positive mass gap. Existence includes establishing axiomatic properties at least as strong as the 
Wightman axioms of  the axiomatic quantum field theory.}  (Slightly edited)
\end{quotation}

Thus the problem is twofold: 
\begin{description}
  \item[A.]  To develop a sufficiently strong mathematically rigorous nontrivial  quantum Yang-Mills theory on the Minkowski space-time.
  \item[B. ]  To  deduce from \emph{that theory}  that there is a positive mass gap in the quantum energy-mass spectrum of Yang-Mills bosons.
\end{description}
A mass gap  for weak and strong forces is suggested by  experiments  in accordance with Hideki Yukawa's principle:  \emph{A  limited force range  indicates a massive carrier}. The heuristic standard model of bosonic particles provides  masses   via a putative classical (i.e., non-quantum) Higgs mechanism. 

\smallskip
Wightman axioms have been formulated in 1950's to  establish a rigorous mathematical framework for  quasi mathematical relativistic quantum field theories of  physicists. 
Not quite "self-evident",  the axioms are inspired    by mathematical properties of free quantum scalar local fields, the operator-valued  solutions of the   linear relativistic Klein-Gordon equation with constant coefficients (see, e.g., \textsc{reed-simon}\cite[Section IX.8]{Reed},  \textsc{strocchi}\cite{Strocchi}). Wightman's   quantum  mass is  the positive  bottom  spectral  gap of a joint unitary representation of the translation subgroup of Poincare symmetries  of the equation. The quantum fields are solutions with values in self-adjoint operators on "physical Hilbert space" of the representation.

The classical relativistic equations of the standard model include relativistic  quasi linear  Yang-Mills  equations for vector fields with components in  gauge compact semi-simple Lie groups. Due to the  additional gauge symmetry, Yang-Mills  equations  are \emph{overdetermined }, a serious challenge even for classical solutions theory. Even more so for Wightman axioms, since the analogous unitary representations  are impossible (see, e.g, \textsc{strocchi}\cite[Appendix A2]{Strocchi}).

\smallskip
Wightman axioms  are  non-dynamical (cp., \cite[Page 215]{Reed}), i.e., neither Lagrangian, nor Hamiltonian. But the concept of mass is dynamical. A relativistic Lagrangian theory of classical  fields  is  described by the  Noether energy-momentum relativistic vector $P^\mu$. Its  \emph{mass} $m>0$  is a relativistic scalar,  provided that
\begin{equation}
\label{eq:sign}
m^2\equiv\ P^\mu P_\mu\ \equiv\ P^0 P^0-P^k P_k  \ \equiv\ P^0 P^0-P^1 P^1-P^2 P^2-P^3 P^3\  > 0,
\end{equation}
i.e.,  $P^\mu$ is a time-like vector.

The   energy-mass  time component  $P^0$  is not a Poincare  scalar and  Einstein's equation $m=P_0$  holds only if the momentum $P^k=0$, i.e., in the distinguished \emph{rest Lorentz frames} where  the  energy-momentum vector is along   the time axis.

The Yang-Mills energy-momentum 4-vector is time-like (see \textsc{glassey-strauss} \cite{Glassey}) in spite of physicists  statement that gauge bosons propagate with the light speed ( by \cite{Glassey},  this holds for the energy-mass density  only asymptotically as $t\rightarrow \infty$). In particular,   the functional  $P_0$  on the Yang-Mills solutions  is  preserved by time translations in   Poincare distinguished frames.  

 This paper presents  a  rigorous  quantization of the  functional $P_0$ in the Yang-Mills distinguished frames and the temporal gauge.

Any  Poincare frame is relativistically equivalent to a  Yang-Mills distinguished frame, and any gauge is equivalent to a temporal gauge. Since the quantization is invariant with respect to the residual Poincare and gauge symmetries, the bosonic spectrum and  its  spectral  gap are Poincare and gauge invariants.
 
The quantization is performed in a Gelfand triple of infinite-dimensional White Noise calculus
(cp. \textsc{hida et al}\cite{Hida} and \textsc{obata}\cite{Obata-94}).
 
The bosonization method (see the companion paper \textsc{dynin}\cite{Dynin-10})
allows further  supersymmetric generalizations.

\subsection{Outline}
\begin{description}
    \item[A1. ]  In the temporal gauge,   Yang-Mills fields (i.e., solutions of \emph{relativistic} Yang-Mills equations) are in one-one correspondence  with their  constrained  Cauchy data. Thus a relativistic Yang-Mills  theory on Minkowski spacetime is equivalent to a Euclidean gauge theory on $\mathbb{R}^3$.

\item[A2. ] This parametrization  of the classical Yang-Mills fields is  advantageous in two ways: 
    \begin{itemize}
  \item The Cauchy data carry a \emph{positive definite} scalar product.  
  \item  The non-linear constraint equation for the Cauchy data is \emph{elliptic}.
\end{itemize}   
The elliptic equation is solved  via a gauge version of  classical Helmholtz decomposition of vector fields.  The solution  provides a global linearization of  the  non-linear constraint  manifold.

\item[A3. ]  In the line of  I. Segal's   quantization program on a space of   Cauchy data (see, e.g.,  \textsc{segal}\cite{Segal-60}) along with   Bogoliubov-Shirkov-Schwinger's prescription   \textsc{bogoliubov-shirkov} \cite[Chapter II]{Bogoliubov}).    The  quantization of the conserved rest energy-mass functional is chosen to be anti-normal (aka anti-Wick  or Berezin  quantization).

  \item[B1. ] Via an infinite-dimensional extension  of  \textsc{agarval-wolf}\cite{Agarwal}'s symbolic calculus we show that the corresponding Weyl  symbol of  the anti-normal energy-mass operator contains a quadratic mass term which is absent in the energy-mass  functional.

  \item[B2. ]    The expectation functional of the  anti-normal energy-mass operator  majorizes  the expectation functional of a shifted   number operator. 
  
   This allows to split off the bosonically irreducible invariant spacess. The corresponding bosonic spectrum is infinite and discrete.
     
   \end{description}

\subsection{Contents}

\quad\ Section 2 reviews  basics of classical Yang-Mills dynamics.

 \smallskip 
 Section 3 describes polynomial operators and their symbols  in Gelfand nuclear triples. 
 
\smallskip
Section 4 defines  bosonic energy-mass spectrum of
Yang-Mills bosons and presents  a proof that it  is infinite, discrete, and grows at least as  an arithmetical progression. 

\smallskip

Section 5  is a sketch of mathematical and physical signposts.

\emph{All new defined terms in the text are introduced via emphasizing in italics.
The beginning and the end  of a proof are marked  by $\triangleright$ and $\triangleleft$.}

\section{Classical   dynamics of  Yang-Mills fields}
\subsection{Gauge groups}

The \emph{global gauge  group}   $\mathbb{G}$ of a  Yang-Mills theory is   a  
connected semi-simple compact Lie group with the  Lie algebra $\mbox{Ad}(\mathbb{G})$. 

The notation $\mbox{Ad}(\mathbb{G})$ indicates that the Lie algebra carries the adjoint representation $\mbox{Ad}(g)X=gXg^{-1}, g\in\mathbb{G}, a\in Ad(\mathbb{G})$, of the group $\mathbb{G}$ and the corresponding self-representation $\mbox{ad}(X)Y=[X,Y],\ X,Y\in\mbox{Ad}(\mathbb{G})$.  Then $\mbox{Ad}(\mathbb{G})$ is identified with a Lie algebra of skew-
symmetric matrices and the matrix commutator as Lie bracket with  the \emph{positive  definite}  Ad-invariant  scalar  product
\begin{equation}
\label{eq:scalar}
X\cdot Y \ \equiv\  \mbox{Trace}(X^TY), 
\end{equation}
where $X^T=-X$ denotes the matrix transposition  (see, e.g., \textsc{zhelobenko}\cite[section 95]{Zhelobenko}).


\bigskip
Let the Minkowski space $\mathbb{M}$ be oriented and time oriented with  the Minkowski metric signature (\ref{eq:sign}). In a Minkowski coordinate systems 
$x^\mu, \mu=0,1,2,3$,  the metric tensor is diagonal.
  In  the natural unit system, the time coordinate $x^0=t$. Thus  $(x^\mu)=(t,x^i),\  i=1,\ 2,\ 3$. 
  
 The \emph{local gauge  group} $\mathcal{G}$ is the group of  infinitely differentiable $\mathbb{G}$-valued functions
 $g(x)$ on $\mathbb{M}$ with the pointwise group multiplication.  The  \emph{local gauge Lie algebra}  $\mbox{Ad}(\mathcal{G})$ consists of  infinitely differentiable $\mbox{Ad}(\mathbb{G})$-valued functions   on $\mathbb{M}$ with the pointwise Lie bracket.   
 
$\mathcal{G}$ acts via the pointwise adjoint action on $\mbox{Ad}(\mathcal{G})$   and correspondingly on  $\mathcal{A}$, the real vector space of \emph{gauge   fields}   $A=A_\mu(x)\in\mbox{Ad}(\mathcal{G})$. 

 \smallskip
   Gauge fields $A$ define   the \emph{covariant partial derivatives}  
   \begin{equation}\label{}
  \partial_{A\mu}X\ \equiv\  \partial_\mu X- \mbox{ad}( A_\mu)X,\quad
X\in\mbox{Ad}(\mathcal{G}).  
\end{equation}
This definition shows that  in the natural units \emph{gauge  connections have the mass dimension} $1/[L]$. 

Any $g\in\mathcal{G}$ defines the affine \emph{gauge transformation}  
\begin{equation}\label{}
A_\mu\mapsto A_\mu^{g}:\ =\ \mbox{Ad}(g)A_\mu-(\partial_\mu g)g^{-1},\ A\in \mathcal{A},
\end{equation}
so that $A^{g_1}A^{g_2}=A^{g_1g_2}$.

\subsection{Yang-Mills fields}
Yang-Mills \emph{curvature tensor} $F(A)$ is the 
antisymmetric tensor
\begin{equation}\label{}
F(A)_{\mu\nu}:\ =\ \partial_\mu A_\nu-\partial_\nu A_\mu-[A_\mu,A_\nu].
\end{equation} 
The curvature is gauge covariant:
 \begin{equation}\label{}
\partial_{A\mu }\mbox{Ad}(g)\ =\ \mbox{Ad}(g)
\partial_{A\mu},\quad
\mbox{Ad}(g)F(A)\ =\ F(A^{g}).
 \end{equation}
 
The \emph{Yang-Mills Lagrangian}
 \begin{equation}\label{}
 \label{eq:Lag}
  L= - (1/4)F(A)^{\mu\nu}\cdot F(A)_{\mu\nu}
  \end{equation}
  is invariant  under gauge transformations.

The corresponding Euler-Lagrange equation is a   2nd order non-linear  partial differential equation $\partial_{A\mu}F(A)^{\mu\nu} =0$, called 
 the \emph{Yang-Mills equation} 
\begin{equation}
\label{eq:YM}
 \partial_\mu F^{\mu\nu}\ +\ [A_\mu, F^{\mu\nu}]\ =\ 0.
\end{equation}
The solutions   $A$ are  \emph{Yang-Mills fields}. They  form the \emph{on-shell space} $\mathcal{M}$ of  the classical Yang-Mills theory. 

\medskip
\emph{From now on we assume  that  all space derivatives of   gauge  fields $A=A(t,x^k)$  vanish faster than any power of $x^kx_k$  as $x^kx_k\rightarrow \infty$, uniformly with respect to bounded $t$}. (This condition does not depend on  a Lorentz coordinate system.) Let $\mbox{Ad}\mathcal{G}$ denote the local Lie algebra of such gauge fields and 
$\mathcal{G}$ denote the corresponding infinite dimensional local Lie group.

\smallskip
Then  3-dimensional integration  of the divergence-free Noether  current vector fields leads to  \emph{Noether  relativistic  and gauge invariant on shell conservation laws}. The 15-dimensional conformal  group of symmetries of Yang-Mills equation produces 15 independent non-trivial  conservation laws (see, e.g., \textsc{glassey-strauss} \cite{Glassey}). Four  of them are the  conservation of the energy-momentum relativistic vector. 

On the other hand, gauge invariance of Yang-Mills equation under infinite dimensional group $\mathcal{G}$ produces no non-trivial   conservation law. In particular, such  Yang-Mills fields are colorless (see, e.g.,  \textsc{glassey-strauss} \cite{Glassey}).

\smallskip
In a Lorentz coordinate system we have the following  matrix-valued  time-dependent fields on  $\mathbb{R}^3$: 
\begin{description}
  \item Gauged electric vector field  $E(A)\equiv(F_{01},F_{02},F_{02})$,
  \item Gauged magnetic pseudo vector field  $B(A)\equiv(F_{23},F_{31},F_{12})$.
  \end{description}
Now the (non-trivial) energy-mass    conservation law   is  that   the time component 
  \begin{equation}
\label{eq:energy}
P^0(A)\equiv\int\!d^3x\, (1/2)(E^i\cdot E_i  + B^i\cdot B_i)
\end{equation}
of the relativistic Noether's  energy-momentum vector is constant on-shell. Appropriately, $P^0(A)$
 has  the mass dimension.

At the same time,  by Glassey-Strauss Theorem \cite{Glassey}, the \emph{energy-mass density}  $(1/2)(E^i\cdot E_i  + B^i\cdot B_i)$    scatters asymptotically along the light cone as $t\rightarrow \infty$. This is a mathematical reformulation of the  physicists assertion that 
Yang-Mills fields propagate with the light velocity.

\subsection{First order formalism}
Rewrite  the    2nd order Yang-Mills equations (\ref{eq:YM})   in the temporal gauge 
$A_0(t,x^k)=0$ as the 1st order systems   of the \emph{evolution equations}  for the time-dependent $A_j(t,x^k)$,  $E_j(t,x^k)$ on  $\mathbb{R}^3$ as
\begin{equation}
\label{eq:evolution}
\partial_t A_k\  = \  E_k ,  \quad
\partial_tE_k \  = \  \partial_jF^j_k - [A_j,F^j_k],\ \quad\ F^j_k\ =\ \partial^j A_k - \partial_k A^j - [A^j,A_k].
\end{equation}
and the \emph{constraint  equations}
\begin{equation}
\label{eq:constraint}
 [A^k,E_k]    \ =\  \partial^kE_k, \quad \mbox{i.e.,}\ \quad \partial_{k,A}E_k\ =\ 0\\
\end{equation}
By  \textsc{goganov-kapitanskii} \cite{Goganov-85}, the evolution system is  a semilinear first order  partial differential  system  with  \emph{finite speed propagation} of the initial data, and the  Cauchy  problem for it with  initial data at $t=0$ \begin{equation}
\label{ }
a(x_k)\ \equiv\ A(0,x_k), \quad \ e(x_k)\ \equiv\ E(0,x_k)
\end{equation}
is \emph{globally and uniquely solvable on} the whole Minkowski space $\mathbb{M}$.

Actually, \textsc{goganov-kapitanskii} proved  this without any restriction on Cauchy data at the  infinity.

\smallskip
As a functional of Cauchy data,  the energy-mass functional (\ref{eq:energy})
is 
\begin{equation}
\label{eq:Lambda}
\Lambda(a,e)\ =\ \int_{\mathbb{R}^3}\!d^3x\: \Big((da - [a,a])\cdot (da - [a,a])\ +\
e\cdot e\Big)
\end{equation}
If the   constraint equations are satisfied  at $t=0$, then, in view of the evolution system, they are satisfied   for  all $t$ automatically. Thus the  \emph{1st order evolution system along with the  constraint equations for Cauchy data is equivalent  to the 2nd order Yang-Mills system}. Moreover the constraint equations are invariant under  \emph{time independent} gauge transformations. As the bottom line, we have
\begin{proposition}
In the temporal gauge Yang-Mills fields $A$ are in one-one correspondence with their  gauge transversal Cauchy data $(a,e)$ satisfying   the  equation $\partial_a e=0$.  
\end{proposition}

Let $\mathcal{A}^0=\mathcal{A}^0(\mathbb{R}^3)$ denote the  \emph{real} $\mathcal{L}^2$-space of Cauchy gauge vector fields  $a$ on $\mathbb{R}^3$.  
The associated   Sobolev-Hilbert spaces  (see, e.g., \textsc{shubin}\cite[Section 25]{Shubin}) are  denoted   $\mathcal{A}^s,\ s\in\mathbb{R}$. The intersection  $\mathcal{A}^\infty_0\equiv\bigcap_s\mathcal{A}^s$  is a nuclear Frechet space of smooth $a$ with the anti-dual union $\mathcal{A}^{-}\infty\equiv\bigcup_s \mathcal{A}^{-s}$.

\smallskip
Let $\mathcal{G}^s,\ s>3/2,$ be the infinite dimensional Frechet Lie groups with the  Lie algebras $\mathcal{A}^s\ s>3/2$.

The intersection $\mathcal{G}^\infty\equiv\bigcap_s\mathcal{G}^s$
is an infinite dimensional Lie group with the \emph{nuclear} Lie algebra
$\mathcal{A}^\infty$. The local gauge transformations $a^g$ by 
$g\in\mathcal{G}^\infty$ define  continuous  left action $\mathcal{G}^\infty\times\mathcal{A}^s\rightarrow
\mathcal{A}^{s-1}$.

  Local gauge transformations 
\begin{equation} \label{eq:gauge}
a_k^g=\mbox{Ad}(g)a_k-(\partial_k g)g^{-1},
\quad g\in\mathcal{G}^\infty,\ a\in \mathcal{A}^s,
\end{equation} 
define continuous  left action of $\mathcal{G}^s$ on $\mathcal{A}^s$.

The Sobolev-Hilbert  spaces  $\mathcal{E}^s$  of  smooth   Cauchy gauge  electric fields $e$ on $\mathbb{R}^3$ with the corresponding action $e^g$  of the local gauge group  $\mathcal{G}^\infty$ are defined the same way.

\smallskip
By  \textsc{dell'antonio-zwanziger} \cite{Dell'Antonio-91},  we have  
 \begin{proposition}
 \label{pr:Gauss}
 Let $\mathcal{G}^0$ denote the completion of  $\mathcal{G}^\infty$ with respect to the natural $\mathcal{L}^2$-metric on the transformations of 
$\mathcal{E}^0$. Then
  \begin{enumerate}
  \item The gauge action of $\mathcal{G}^\infty$  on $\mathcal{A}^\infty\times \mathcal{E}^\infty$  has a unique extension to the continuous action of $\mathcal{G}^0$ on
  \begin{equation}
\label{ }
\mathcal{C}^0\ \equiv\  \mathcal{A}^0\times \mathcal{E}^0 .
\end{equation}
  
  \item  The gauge orbits of this action are closures of  $\mathcal{G}^\infty$-orbits.
  
    \item On the  orbit of every $e$ the Hilbert $\mathcal{L}^2$-norm $\|a^g\|$ attains the absolute minimum  at some gauge  equivalent connection 
    $\breve{a}\in\mathcal{A}^0$. 
   \item Minimizing connections $\breve{a}$ are weakly divergence free:
$\partial^k\breve{a}_k\ =\ 0$. 
\end{enumerate}
\end{proposition}

\subsection{Gauged vector calculus}

Let $\mathcal{U}^s$  denote the Sobolev-Hilbert spaces $\mbox{Ad}(\mathbb{G})$-valued functions $u$ on $\mathbb{R}^3$.

Consider the continuous  vector calculus  operators gauged by $a\in\mathcal{A}^{\infty}$

\smallskip
\emph{Gauged gradient}
 \begin{equation}
\label{ }
  \mbox{grad}_{a}:\ \mathcal{U}^{s}\rightarrow \mathcal{E}^{s-1},\quad\ \mbox{grad}_{a}u \ \equiv  \partial_k u - [a_k,u],
\end{equation} 
\indent\emph{Gauged divergence}
\begin{equation}
\label{ }
  \mbox{div}_{a}:\ \mathcal{E}^{s}\rightarrow \mathcal{U}^{s-1},\quad\ 
  \mbox{div}_{a}e \ \equiv  \partial_k e_k - [a_k,e_k],
\end{equation} 
\indent\emph{ Gauged Laplacian}
\begin{equation}
\label{ }
  \triangle_a :\ \mathcal{U}^{s}\rightarrow \mathcal{U}^{s-2},\quad\  
  \triangle_a \ \equiv\  \mbox{div}_{a}u\,\mbox{grad}_{a}u,
\end{equation} 
 The 1st order partial differential operators $-\mbox{grad}_{a}$ and  $\mbox{div}_{a}$  are adjoint  with respect to the $\mathcal{L}^2$ scalar product:
\begin{equation}\label{adjoint}
\langle\  -\mbox{grad}_{a} u\ |\ v\ \rangle\ =\  
\langle\, u\ |\ \mbox{div}_{a}v\, \rangle.
\end{equation} 
The  gauge Laplacian $\triangle_a$ is a  2nd order partial differential operator. Since its principal part is the usual Laplacian $\triangle$, the operator $\triangle_a$ is elliptic. \begin{proposition}
\label{pro:Lap}
The gauge Laplacian  $\triangle_a$ is  an invertible operator  from 
$\mathcal{U}^{s+2}$ onto $\mathcal{U}^{s}$ for all $s\geq 0$.
\end{proposition}
\begin{lemma}\label{lemma:inj}
$\triangle_a u=0,\ u\in\mathcal{U}^1_0,$ if and only if  $u=0$.
\end{lemma}  
\proof
 $u\cdot [a,u]=-\mbox{Trace}(uau-uua)=0$ so that
\begin{equation}
\label{ } 
u\cdot \mbox{grad}_{a}\,u \ =
u\cdot\mbox{grad}\, u \ =\  
(1/2)\mbox{grad}\,(u\cdot u)\ =\ 0.
\end{equation}
This shows that   for $u\in\mathcal{U}^1_0$ we have $\mbox{grad}_a u =0$    if and only if  $u=0$. \qed

Next, by the equality (\ref{adjoint}),
\begin{equation}\label{}
\langle\, \triangle_a u\ |\ u\, \rangle\ =\
\langle\  -\mbox{grad}_{a}u\ |\ \mbox{grad}_{a}u\ \rangle,\   u\in\mathcal{U}^1_0.
\end{equation}
Thus   $\triangle_a u=0,\ u\in\mathcal{U}^1_0,$ if and only if  $u=0$. \qed

\smallskip
Both Laplacian  $\triangle$ and gauge Laplacian    $\triangle_a$ map 
$\mathcal{U}^{s+2}$ into $\mathcal{U}^s$.

The Laplace operator 
 is invertible from  $\mathcal{U}^{s+2}$ onto $\mathcal{U}^{s}$ whatever $s\geq 0$ is.
Since $\triangle - \triangle_a$ is a 1st order differential operator, the operator 
 $\triangle_a:\mathcal{U}^{s+2}\rightarrow\mathcal{U}^s$ is a Fredholm operator of zero index. Then, by   Lemma \ref{lemma:inj}, the inverse 
$\triangle_{a}^{-1}:\ \mathcal{U}^{s}\rightarrow\mathcal{U}^{s+2}$ exists
for all $s\geq 0$. \qed

Now proposition \ref{pro:Lap} shows that the operator  $\mbox{div}_a: \mathcal{U}^s\rightarrow\mathcal{U}^{s-1}$ is surjective and
 the operator  $\mbox{grad}_a: \mathcal{U}^s\rightarrow\mathcal{U}^{s-1}$ is injective. Therefore,\begin{theorem}
\label{pro:div}
The gauged Helmholtz operator 
\begin{equation}\label{Helm}
P_a\ \equiv\ 
 \mbox{grad}_a\triangle_{a}^{-1}\mbox{div}_{a}
\end{equation}
is an  $\mathcal{L}^2$-orthogonal projector of $\mathcal{U}^s$ onto  the space of gauge longitudinal  vector fields, i.e., the range   of the operator 
$\mbox{grad}_a:\ \mathcal{U}^{s+1}\rightarrow \mathcal{U}^s$.

The operator $\mathbf{1}-P_a$ is an $\mathcal{L}^2$ bounded projector of $\mathcal{U}^s$ onto  the space of gauge transversal vector fields, i.e., the null space  of the operator 
$\mbox{div}_{a}: \mathcal{U}^s\rightarrow\mathcal{U}^{s-1}$.
\end{theorem}
\proof   
Both $P_a$ and $\mathbf{1}-P_a$ are  pseudodifferential operators of order 0, and,
therefore are $\mathcal{L}^2$- bounded.

By computation, 
\begin{equation*}
\label{ }
P_a^\dag=P_a,\quad P_a^2=P_a,\quad P_a\mbox{grad}_{a}=\mbox{grad}_{a},\quad 
\mbox{div}_a(\mathbf{1}-P_a)=0. 
\end{equation*}

\section{Yang-Mills bosonic spectrum} 

\subsection{Gelfand triple of consraints}

 
Let  $\mathcal{T}^\infty_a\subset\mathcal{E}^\infty$  denote  the nuclear  Frechet  space  of gauge transversal  gauge electric vector fields $e_a\equiv \ e-P_a(e)$, and $\mathcal{T}^0$ be its completion in  $\mathcal{E}^0$.

The family of orthogonal projectors  $a\mapsto P_a$ is a continuous mapping of  
 $\mathcal{A}^\infty$ to the algebra of bounded operators on $\mathcal{E}^0$.  Since for $a$ sufficiently close to $a_o$ the  operators $1-P_a+P_{a_o}$ are invertible and
$P_aP_{a_o}=P_a(1-P_a+P_{a_o}))P_{a_o}$,
the continuous mappings $P_aP_{a_o}:  P_{a_o}(\mathcal{E}^0)\rightarrow  P_{a}(\mathcal{E}^0)$
are invertible. Thus  the  vector  bundle $\mathcal{T}^0$ of the gauge transversal  spaces 
  $\mathcal{T}^0_a$ is  a locally trivial   real   vector  bundle over  $\mathcal{A}^\infty$.
  
  Since the projectors  $P_a$ are pseudodifferential operators, the vector bundle $\mathcal{T}^\infty$ of $\mathcal{T}^\infty_a$ is a locally trivial the bundle  over   $\mathcal{A}^\infty$.

\smallskip
Gauge invariance of the constraint manifold of Cauchy data under the (residual) gauge group  implies the gauge covariance of projectors $\mathbf{1}-P_a$, and so of  the  bundles.
Since   a  Hilbert  bundle structure group is   smoothly contractible  (see \textsc{kuiper}\cite{Kuiper}), the bundle $\mathcal{T}^0$
is isomorphic to the trivial gauge covariant Hilbert space bundle over its base:  an isomorphism is defined by a smooth family of orthonormal bases of the bundle fibers. All such trivialisations intertwine with the action of the residual gauge group.
They define linearly isomorphic global Hilbert  coordinate charts on the constraint Cauchy data manifold $\mathcal{C}^0\ \cong \mathcal{A}^0\times \mathcal{T}^0_0$ along with    the natural Gelfand nuclear triple of real topological vector spaces
\begin{equation}\label{eq:Gelfand}
\mathcal{C}:\ \mathcal{C}^\infty\  \equiv \mathcal{A}^\infty_0\times 
\mathcal{T}^\infty_0\subset\ \mathcal{C}^0\ \equiv \mathcal{A}^0\times 
\mathcal{T}^0\subset\ \mathcal{C}^{-\infty}\ \equiv \mathcal{A}_0^{-\infty}\times 
\mathcal{T}_0^{-\infty}.
\end{equation}
where $\mathcal{C}^\infty$  is a nuclear Frechet space of smooth $(a,e^o)$, and 
$\mathcal{C}_0^{-\infty}$ is  the  dual of $\mathcal{C}^\infty_0$,  with the duality defined by the inner product in $\mathcal{C}^0$.

\medskip
The assignment $(a,e^o)\mapsto z=(1/\sqrt{2})(a+ie^o)$ converts the real  Gelfand triple (\ref{eq:Gelfand})
into the complex  Gelfand triple
\begin{equation}
\label{eq:CGelfand}
\mathcal{C}_{\mathbb{C}}:\ \mathcal{C}_{\mathbb{C}}^\infty\ \subset\ \mathcal{C}_{\mathbb{C}}^0\ \subset\ 
\mathcal{C}_{\mathbb{C}}^{-\infty},
\end{equation}
so that $\Re\mathcal{C}_{\mathbb{C}}\ \equiv\mathcal{A}$ and 
$\Im\mathcal{C}_{\mathbb{C}}\ \equiv\mathcal{T}_o$ are its real and imaginary parts.

 The  complex conjugation
\begin{equation}
\label{ }
z^*\ =\  (1/\sqrt{2})(a+ie^o)^*\ \equiv\ (1/\sqrt{2})(a-ie^o),\ z\mapsto z^*:\ \mathcal{C}_{\mathbb{C}}\rightarrow
\mathcal{C}_{\mathbb{C}}^{-\infty}
\end{equation} 
The  (anti-linear on the left and linear on the right) Hermitian form $z^*w$  defined  on  by
 $\mathcal{C}_{\mathbb{C}}^0$
\begin{equation}
\label{eq:square}
z^*z\ \equiv\ (1/2)\int\!d^3 x\ (a\cdot a + e^o\cdot e^o)
\end{equation}
is extended to the anti-duality between $\mathcal{C}_{\mathbb{C}}^\infty$ and , 
$\mathcal{C}_{\mathbb{C}}^{-\infty}$. Accordingly,  \emph{the notation $z$ is reserved for the elements of the former space , and  the notation $z^*$ for the elements of the latter space}.

\subsection{Quantization}
The nuclear Gelfand triple $\mathcal{C}_{\mathbb{C}}$ is a standard  Hida triple of White Noise calculus (cp.  \textsc{hida et al}\cite{Hida}). Its  canonical quantization 
(see, e.g.,  \textsc{obata}\cite{Obata-94}) is a Gelfand triple with complex conjugation
\begin{equation}
\label{ }
(\mathcal{C}_{\mathbb{C}}):\ (\mathcal{C}_{\mathbb{C}})^{\infty}\
\subset\ (\mathcal{C}_{\mathbb{C}})^0\ \subset\ (\mathcal{C}_{\mathbb{C}})^{-\infty}.
\end{equation}
 carrying the  \emph{canonical representation} of   $\mathcal{C}_{\mathbb{C}}$  by continuous linear transformations of $z$ and $z^*$ into  adjoint  linear operators of creation and annihilation
\begin{eqnarray}
& &
\hat{z}:\ (\mathcal{C}_{\mathbb{C}})^{\infty}\ \rightarrow (\mathcal{C}_{\mathbb{C}})^{\infty},\quad
\widehat{z^*}:\ (\mathcal{C}_{\mathbb{C}})^{-\infty}\ \rightarrow (\mathcal{C}_{\mathbb{C}})^{-\infty},\\
& &
\widehat{z^*}^\dag:\ (\mathcal{C}_{\mathbb{C}})^{\infty}\ \rightarrow (\mathcal{C}_{\mathbb{C}})^{\infty},\quad
\hat{z}^\dag:\ (\mathcal{C}_{\mathbb{C}})^{-\infty}\ \rightarrow (\mathcal{C}_{\mathbb{C}})^{-\infty},
\end{eqnarray} 
assuming
\begin{enumerate}
  \item 
 Bosonic commutation  relation
\begin{equation}
\label{eq:CCR}
[\widehat{\zeta^*}^\dag,\hat{z}]\ =\ \zeta^*z.
\end{equation}
  \item Existence of a unique unit fiducial real state
$\Omega_0\in(\mathcal{C}_{\mathbb{C}})^\infty$ (aka \emph{vacuum state}) such that
\begin{equation}
\label{ }
 \hat{z}^\dag\Omega_0\ = 0\ =\ \widehat{z^*}^\dag\Omega_0.
\end{equation}
\item 
The set of the \emph{coherent states}
\begin{equation}
\Omega_z\ \equiv\  \sum_{n=0}^\infty (1/n!)\hat{z}^n\Omega_0\ \in\ (\mathcal{C}_{\mathbb{C}})^\infty
\end{equation}
is total, i.e., if $\Psi^*\Omega_z=0$ for all $\Omega_z$
then $\Psi=0$.  Furthermore, 
\begin{equation}
\label{eq:Gauss}
\Omega_{\zeta}^*\Omega_z=e^{\zeta^*z}
\end{equation}
\end{enumerate}
The  S-\emph{transforms} (\textsc{obata}\cite{Obata-94}) (cp. generating functionals \textsc{berezin}\cite{Berezin-65}) of $\Psi^*\in(\mathcal{C}_{\mathbb{C}})^{-\infty}$ and
$\Psi\in(\mathcal{C}_{\mathbb{C}})^\infty$  and 
\begin{equation}
\label{eq:S} 
\Psi^*(z)\ \equiv\ \Psi^*\Omega_z,\quad \Psi(z^*))\ \equiv\ (\Omega_z)^*\Psi
\end{equation}
are entire functionals  correspondingly on $\mathcal{C}_{\mathbb{C}}^{\infty}$ and 
$\mathcal{C}_{\mathbb{C}}^{-\infty}$. 

By the characterization theorems (see, e.  g., \textsc{obata},  \cite[Theorems 3.7 and 3.6]{Obata-94}) S-transform is a topological linear isomorphism of 
$(\mathcal{C}_{\mathbb{C}}^{\infty})$ onto the topological algebra with the point-wise multiplication of entire functionals  of the bornological order 2 and type 0 on 
$\mathcal{C}_{\mathbb{C}}^{-\infty}$, as well as  a topological linear isomorphism of 
$(\mathcal{C}_{\mathbb{C}}^{-\infty})$ onto the topological algebra with the point-wise multiplication of entire functionals  of the topological order 2  on 
$\mathcal{C}_{\mathbb{C}}^{\infty}$.\footnote{Interpretation is mine. A.D.}

The identities  
\begin{equation}
\label{ } 
\hat{\zeta}^\dag\Omega_z^*\  \stackrel{(\ref{eq:CCR})}{=}\ (\zeta^*z)\Omega_z^*,\quad \hat{\zeta}\Omega_z^*\  \stackrel{(\ref{eq:Gauss})}{=}\ \partial_{\zeta}\Omega_z^*,
\end{equation}
imply, by (\ref{eq:S}), their adjoints
\begin{equation}
\label{eq:inter}
\hat{\zeta}\Psi(z^*) =\ (z^*\zeta)\Psi(z^*),\quad \hat{\zeta}^\dag\Psi(z^*)\ =\  \partial_{\zeta^*}\Psi(z^*).
\end{equation}

\smallskip
Henceforth we  use Einstein's convention for tensor  contraction along \emph{conjugated  continual indices}:
 \begin{equation}
  \Phi^*(z)\Psi(z^*)\ \equiv\ \Phi^*\Psi.
  \end{equation}

\medskip
The bosonic quantization of the direct product 
\begin{equation}\label{eq:product}
\mathfrak{C}:\ \mathcal{C}_{\overline{\mathbb{C}}}^\infty\times \mathcal{C}_{\mathbb{C}}^\infty\ \subset\ \mathcal{C}_{\overline{\mathbb{C}}}^0  \times \mathcal{C}_{\mathbb{C}}^0\ \subset\ 
\mathcal{C}_{\overline{\mathbb{C}}}^{-\infty}\times \mathcal{C}_{\mathbb{C}}^{-\infty}
\end{equation}
with  the complex conjugation $(z^*,w)^*\equiv (w^*,z)$ produces 
the sesquilinear Gelfand triple $(\mathfrak{C})$. The corresponding  coherent states are
\begin{equation}
\label{ }
\Omega_{(z^*,w)}\ =\ \Omega_z \Omega_{w^*}.
\end{equation}


\subsection{Operator symbols}
Creators and annihilators generate strongly continuous abelian operator groups in 
$(\mathcal{C}_{\mathbb{C}})^{\infty}$ and $(\mathcal{C}_{\mathbb{C}})^{-\infty}$ parametrized by $\zeta$ and $\zeta^*$:
\begin{eqnarray}
\label{eq:1}
e^{\hat{\zeta}}: (\mathcal{C}_{\mathbb{C}})^{\infty}\rightarrow (\mathcal{C}_{\mathbb{C}})^{\infty},\quad & &
e^{\hat{\zeta}}\Psi(z^*)\ =\ e^{\zeta}\ \Psi(z^*);\\
\label{eq:2}
e^{\hat{\zeta}^\dagger}: (\mathcal{C}_{\mathbb{C}})^{\infty}\rightarrow (\mathcal{C}_{\mathbb{C}})^{\infty},\quad & &
e^{\hat{\zeta}^\dagger}\Psi(z^*)\ =\Psi(\zeta^*+z^*);\\
\label{eq:3}
e^{\widehat{\zeta^*}}: (\mathcal{C}_{\mathbb{C}})^{-\infty}\rightarrow (\mathcal{C}_{\mathbb{C}})^{-\infty},\quad & &
e^{\widehat{\zeta^*}}\Psi(z)\ =\ e^{\zeta^*}\ \Psi(z);\\
\label{eq:4}
e^{\widehat{\zeta^*}^\dagger}: (\mathcal{C}_{\mathbb{C}})^{-\infty}\rightarrow (\mathcal{C}_{\mathbb{C}})^{-\infty},\quad & &
e^{\widehat{\zeta^*}^\dagger}\Psi(z)\ =\  \Psi(z+\zeta).
\end{eqnarray}
 Baker-Campbell-Hausdorff commutator formula entails from (\ref{eq:CCR})
 \begin{equation}
\label{eq:BCH}
e^{\hat{\zeta} +\hat{\zeta}^\dag}\ =\ 
 e^{-\zeta^*\zeta/2}e^{\hat{\zeta}^\dag} e^{\hat{\zeta}} :\ (\mathcal{C}_{\mathbb{C}})^{\infty}\rightarrow (\mathcal{C}_{\mathbb{C}})^{\infty}.
\end{equation}
Sesquientire functionals  $\Theta(\zeta^*,\eta)\in(\mathfrak{C})^{-\infty}$ are uniquely 
 defined  by   their restrictions  $\Theta(\zeta^*,\zeta)$ to the real part $\Re\mathfrak{C}^\infty$ of $\mathfrak{C}^\infty$.  
 
  \emph{Normal, Weyl,anti-normal quantizations}  of sesquientire functionals 
   are   the continuous linear operators from  $(\mathcal{C}_{\mathbb{C}})^{\infty}$ to
   $(\mathcal{C}_{\mathbb{C}})^{-\infty}$ defined (in the  continual Einstein's contraction notation   over
    $\Re\mathfrak{C}^\infty$) as
 \begin{equation}
\label{eq:quantizations}
\widehat{\Theta}_n\equiv\ \Theta_n(\zeta^*,\zeta)e^{\hat{\zeta}}e^{\hat{\zeta}^\dagger}, 
\widehat{\Theta}_w\equiv\ \Theta_w(\zeta^*,\zeta)e^{\hat{\zeta} + \hat{\zeta}^\dagger},
\widehat{\Theta}_{a}\equiv\ \Theta_{an}(\zeta^*,\zeta)e^{\hat{\zeta}^\dagger}e^{\hat{\zeta}}.
\end{equation}
The  coherent states matrix element of $\widehat{\Theta}_n$ at $ \Omega_v= \Omega_v(z^*),\ \Omega_u=\Omega_u(z^*)$
  \begin{eqnarray}
\label{eq:matrix}
& &
( \Omega_v)^*\widehat{\Theta}_n \Omega_u\ = \ 
( \Omega_v)^*\Theta_n(\zeta^*,\zeta)e^{\hat{\zeta}}e^{\hat{\zeta}^\dagger}\Omega_u\\
& &
 =\ (e^{\hat{\zeta}^\dagger} \Omega_v)^*\Theta_n(\zeta^*,\zeta)
 (e^{\hat{\zeta}^\dagger}e^u) \\
 & &
=\ (e^{\zeta^*w} \Omega_v)^*\ \Theta_n(\zeta^*,\zeta)(e^{\zeta^*u}\Omega_u) \\
 & &
 =\  ( \Omega_v)^*(e^{v^*\zeta}\Theta_n (\zeta^*,\zeta)e^{\zeta^*u}\Omega_u)\\
 & &
 =\   \Theta_n(u^*,v)e^{v^*u},
\end{eqnarray}
where $\Theta_n(u^*,v)$ is the   S-transform of $\Theta_n(\zeta^*,\zeta)$.
Its   restriction $\Theta_n(z^*,z)$ to $\Re\mathfrak{C}^\infty$  is the \emph{normal symbol} of  the operator $\widehat{\Theta}_n$. 

Any continuous  linear operator $Q$ from $(\mathcal{C}_{\mathbb{C}})^{\infty}$ to $(\mathcal{C}_{\mathbb{C}})^{-\infty}$ is, by Grothendieck kernel theorem, the normal quantization of a unique $\Theta^Q_n(z^*,z)\in\Re(\mathfrak{C})^{-\infty}$. Then, by (\ref{eq:BCH}),  $Q$ is also the Weyl and anti-normal quantizations  of  unique classical variables  
$\Theta^Q_w(z^*,z)$ and  $\Theta^Q_{an}(z^*,z)$. 
Their S-transforms  $\Theta^Q_w(\zeta^*,\zeta)$ and $\Theta^Q_{an}(\zeta^*,\zeta)$ are  the \emph{Weyl} and  \emph{antinormal symbols} of  the operator $Q$.
 
Moreover, by (\ref{eq:BCH}), (\ref{eq:quantizations}), and (\ref{eq:inter}),
we get infinite-dimensional versions  of Weierstrass transforms
(cp.\textsc{agarwal-wolf}\cite[formulas (5.29), (5.30), (5.31), page 2173]{Agarwal}):
\begin{eqnarray}
\label{eq:wn}
\Theta^Q_w(\zeta^*,\zeta) & = &  e^{-(1/2)\partial_\zeta\partial_{\zeta^*}}\Theta^Q_n(\zeta^*,\zeta),\\
\label{eq:an}
\Theta^Q_{an}(\zeta^*,\zeta) & = & e^{-\partial_\zeta\partial_{\zeta^*}}\Theta^Q_n(\zeta^*,\zeta),\\
\label{eq:wa}
\Theta^Q_w(\zeta^*,\zeta) & = &  e^{(1/2)\partial_\zeta\partial_{\zeta^*}}\Theta^Q_{an}(\zeta^*,\zeta),
\end{eqnarray}
where the Laplacian $\Delta\equiv\partial_\zeta\partial_{\zeta^*}$ is the S-transform of the  multiplication operator  $\Theta^Q(z^*,z)\mapsto (z^*z)\Theta^Q(z^*,z)$. 

Since $z=(1/\sqrt{2})(a+ie^o)$, this is the continuous Gross-Laplace  operator
(see, e.g, \textsc{obata}\cite[Section 5.3]{Obata-94})
\begin{equation}
\label{eq:Gross}
\Delta\ =\ \partial_a^2\ +\ \partial_{e^o}^2: 
(\mathcal{C}_{\mathbb{C}})^{\infty}\rightarrow(\mathcal{C}_{\mathbb{C}})^{\infty}.
\end{equation}
By (\ref{eq:BCH}) and (\ref{eq:matrix}), the   matrix element of
$\widehat{\Theta}_{an}$
 \begin{eqnarray}
& &
 \label{eq:antiWick}
 (\Phi)^*\widehat{\Theta}_{an}\Psi\ \ = \ 
(\Psi)^*\Theta_{an}(z^*,z)e^{\hat{z}^\dagger}e^{\hat{z}}\Psi\\
& &
=\ (e^{\hat{z}}\Phi)^*\Theta_{an}(z^*,z)(e^{\hat{z}}\Psi\
=\ (e^{z^*z}\Psi)^*\Theta_{an}(z^*,z)(e^{z^*z}\Psi)\\ 
& &
 =\ (\Phi)^*(e^{z^*z} \Theta_{an} (z^*,z)e^{z^*z})\Psi\\ 
& &
 =\ (\Phi)^* \Theta_{an}(z^*,z)\Psi\ =\ \Theta_{an}(z^*,z) (\Phi)^*(z)\Psi(z^*).
\end{eqnarray}
The latter implies
\begin{proposition}
The expectation
\begin{equation}
\label{eq:geq}
\langle\widehat{\Theta}_{an}\rangle\ \geq\ \inf \ \Theta_{an}(z^*,z).
\end{equation}
\end{proposition}

An operator $Q$ is  a \emph{polynomial operator}  if its normal symbol (and then the other symbols) is a continuous polynomial  on $\mathcal{C}_{\mathbb{C}}^*\times \mathcal{C}_{\mathbb{C}}$, and, in particular, belongs to $\mathfrak{B}^\infty$.

Equations (\ref{eq:1}) -- (\ref{eq:4}) imply (cp. \textsc{obata}\cite[Section 4.4]{Obata-94})
\begin{proposition}
Polynomial operators are continuous linear transformations of $(\mathcal{C}_{\mathbb{C}})^{\infty}$ into itself
and of $(\mathcal{C}_{\mathbb{C}})^{-\infty}$ into itself.

Furthermore, their operator products of polynomial operators are polynomial.
\end{proposition}

The \emph{number operator}  $N\equiv\hat{z}\widehat{z^*}^\dagger:\ (\mathcal{C}_{\mathbb{C}})^{\infty}\rightarrow (\mathcal{C}_{\mathbb{C}})^{\infty}$  and $(\mathcal{C}_{\mathbb{C}})^{-\infty}\rightarrow (\mathcal{C}_{\mathbb{C}})^{-\infty}$ has  the  symbols 
\begin{equation}
\label{eq:number}
\Omega^N_n(z^*,z)= zz^*,\ \Omega^N_w(z^*,z)= zz^*+1/2,\ \
\Omega^N_{an}(z^*,z)=zz^*+ 1.
\end{equation}
The eigenspaces $\mathcal{N}_n,\   n=0,1,2,...,$ of $N$  with the corresponding eigenvalue $n$ is the  space of continuous  homogeneous polynomials of degree ($n$-\emph{bosonic states}). 

In particular,  the constant  \emph{vacuum state}  $\Psi_0\equiv 1$ corresponds to the eigenvalue $n=0$. In general, homogeneous polynomials of degree $n$ on a complex vector space  are functionals whose restrictions to finite dimensional complex vector subspaces are finite dimensional  homogeneous polynomials of degree $n$. 

\smallskip
The triple $(\mathcal{C}_{\mathbb{C}})$ is the topological orthogonal sum of $n$-bosonic Gelfand triples
\begin{equation}
\label{eq:n}
\mathcal{N}_n^\infty\ \subset\ \mathcal{N}_n^0\ \subset 
\mathcal{N}_n^{-\infty},\quad n=0,1,2,....
\end{equation}

\subsection{Energy-mass bosonic spectrum of Yang-Mills bosons} 

The energy-mass  functional (\ref{eq:Lambda}) on smooth  transversal Cauchy data with minimizing
 $\breve{a}$  
 \begin{eqnarray}
& &
\Lambda(\breve{a},e)\ =\ \int_{\mathbb{R}^3}\!d^3x\: \Big((d\breve{a} - [\breve{a},\breve{a}])\cdot (d\breve{a} - [\breve{a},\breve{a}])\ +\
e\cdot e\Big)\\
& &
\label{eq:short}
=\ \int_\mathbb{R}^3\!d^3x\: \Big(d \breve{a} \cdot d\breve{a}\ + 
 [\breve{a},\breve{a}]\cdot[\breve{a},\breve{a}]\ +\ e\cdot e\Big)
 \end{eqnarray}
has no cubic terms  since, by Proposition \ref{pr:Gauss}, the minimizing connections 
$\breve{a}$ are divergence free. Thus, by gauge invariance, the energy-mass  functional is positive.

\emph{Henceforth we deal  only with minimizing connections removing the "breves" from the notation.}

\smallskip
Let the  \emph{quantum Yang-Mills energy-mass operator} $H:\ (\mathcal{C}_{\mathbb{C}})^\infty
\rightarrow (\mathcal{C}_{\mathbb{C}})^{\infty}$  be the anti-normal quantization  of the   energy-mass functional  $\Lambda$:
\begin{equation}
\label{eq:H}
\Theta^H_{an}(a,e)\ \equiv\ \Lambda(a,e)\ =\ \Lambda(z^*,z  ),\quad  \zeta=a+ie,\ \zeta^*=a^T-ie^T,
\end{equation}
i.e., $\Lambda$ is  the \emph{anti-normal symbol}  of $H$.

\smallskip
The \emph{expectation functional} on non-zero $(\mathcal{C}_{\mathbb{C}})^\infty\in(\mathcal{C}_{\mathbb{C}})^\infty$ of a polynomial   operator $Q$ is 
\begin{equation}
\label{ }
\langle Q \rangle (\Psi) \equiv\ \Psi^*Q\Psi/\Psi^*\Psi.
\end{equation}
  
\begin{proposition} \label{pr:geq}
There exists a constant scalar  field $C$ on $\mathbb{R}^3$ such that 
the expectation functional     
\begin{equation}
\label{}
\langle H\rangle \ \geq \ \langle N\rangle  \  + \ \langle C\rangle,
\end{equation}
where $N$ is the number operator (\ref{eq:number}). 
\end{proposition}

\proof 

(A)\  Let $M$ be the  operator with the non-negative anti-normal symbol 
\begin{equation}
\label{eq:M}
\Omega^M_{an}(z^*,z)\ \equiv\ \int_{\mathbb{R}^3}\!d^3x\:([a,a]\cdot [a,a]+ e\cdot e).
\end{equation}
Then, by (\ref{eq:geq}) and (\ref{eq:short}). 
\begin{equation}
\label{eq:A}
\langle H\rangle \ \geq \ \langle M\rangle.
\end{equation}

(B)\ Let  $b_i$ be a basis for $\mbox{Ad}(\mathbb{G})$ with $b_i\cdot b_j=\delta_{ij}$.
Then the structure constants $c^k_{ij}\ =\ [b_i,b_j]\cdot b_k$
 are anti-symmetric under interchanges of  $i,j,k$.  
Thus if   $a=a^ib_i\in\mbox{Ad}(\mathbb{G})$ then 
\begin{equation}
\label{eq:Killing}
a\cdot a\ =\  \mbox{Trace}(a^ta)\ =\ -a^ic^k_{ij}a^lc^j_{kj}\ =\  a^ic^k_{ij}a^lc^k_{lj},
\end{equation}
so that
\begin{eqnarray}
& & 
[a,a]\cdot[a,a]\ =\  a^ia^ja^la^m\, [b_i,b_j]\cdot  [b_l,b_m]\\
& & 
\label{eq:square}
=\  a^ia^ja^la^m c^k_{ij}c^k_{lj}\ =\  \sum_k(a^ia^jc^k_{ij})^2,
\end{eqnarray}
and the Gross Laplacian  (\ref{eq:Gross})
\begin{equation}
\label{eq:Laplacian}
\Delta([a,a]\cdot[a,a])\ \stackrel{(\ref{eq:Killing})}{=}\ 2a\cdot a.
\end{equation}
Then there is a constant scalar field $C$ such that the Weyl symbol of the operator $M$
\begin{equation}
\label{ }
\Theta^M_w(a,e)\ \stackrel{(\ref{eq:wa}),(\ref{eq:Laplacian})}{=}\ 
\int_{\mathbb{R}^3}\!d^3x\:\big([a,a]\cdot [a,a]\ +\ a\cdot a \ +\ e\cdot e\big)\ +\ 1/2\ +\ C.
\end{equation}
(C)\ The Weyl quantization of  $[a,a]\cdot [a,a]$  is the operator of multiplication with $[a,a]\cdot [a,a]\geq 0$ in the "$(a,e)$-representation" of the canonical commutation relations (cp. \textsc{agarwal-wolf}\cite[Section VII, page 2177]{Agarwal}). In paricular,  its expectation functional is non-negative.

(D)\ By (\ref{eq:number}),  $\int_{\mathbb{R}^3}\!d^3x\:(a\cdot a \ +\ e\cdot e)\ +\ 1/2$   is the anti-normal symbol of the number operator $N$.

Thus 
\begin{equation}
\label{eq:B}
\langle M \rangle\  \geq \langle N \rangle\ +\ \langle C \rangle.
\end{equation} 
The propostion  follows from  the inequalities (\ref{eq:A}) and (\ref{eq:B}). \qed

\bigskip
Operfator   $H$ is a polynomial symmetric operator with non-negative expectation functional. By \textsc{obata}\cite[Proposition 4.5.5]{Obata-94}, a polynomial  operator cannot be  bounded on $(\mathcal{C}_{\mathbb{C}})^0$. However, $H$ has  a  unique  Friedrichs  extension to an unbounded self-adjoint  operator on that Hilbert space.  Proposition \ref{pr:geq} implies via the variational mini-max principle (see, e.g., \textsc{berezin-shubin}\cite[Appendix 2,Proposition 3.2]{Berezin-91}) that its  spectrum is degenerate along with the spectrum of the number operator.

To remove the degeneracy, consider the $n$-particle spaces $\mathcal{N}^\infty_n$ as elementary bosons of spin $n$. Then define the
\emph{bosonic spectrum} of $H$ as  the non-decreasing sequence of its   \emph{spectral values} 
 \begin{equation}
\label{eq:variational}
 \lambda_n(H)\equiv \inf\{\langle H \rangle(\Psi),\ \Psi\in \mathcal{N}^\infty_n\}.
 \end{equation}
 Proposition \ref{pr:geq} implies the enhanced titular 
\begin{theorem}
\label{th:titular} 
The bosonic spectrum of Yang-Mills energy-mass  operator $H$ is infinite and discrete, i.e., each 
 $\lambda_n(H)$  has a  finite multiplicity.

The spectral values  grow at least in  the arithmetical progression:   
\begin{equation}
\label{}
\lambda_n(H)\ \geq\ n\ + \mbox{constant},\quad n=0,\ 1,\ 2, \dots\ .
\end{equation}
\end{theorem}

\section{Signposts}

\subsection{Beyond Hilbert spaces}
Almost immediately after W. Heisenberg's  (1925)  and  E. Shr\"{o}dinger's (1926) formulations of quantum mechanics  von Neumann (1932) and Weyl (1931) created corresponding  new mathematics. 

Von Neumann  defined and named Hilbert spaces to honor Hilbert theory of quadratic forms. He replaced the latter by  (unbounded) self-adjoint operators on such spaces to represent   quantum observables.  
 
 Independently, Weyl's    quantization rule converts  classical  observables into partial differential operators.  Following up    quantization rules proposed by R. Glauber,  E. Sudarshan, et al,  along with their  formal calculus, were generalized  by \textsc{agarwal-wolf}\cite{Agarwal}.
Mathematically, the calculus is related to  theory of pseudodifferential operators. (see, e.g., \textsc{shubin}\cite[Chapter 4]{Shubin}).

Even up to this day quantum field theory  has remained mathematically challenged.
Actually, P. Dirac was not fascinated with Hilbert spaces preferring his own bra-ket formalism inspired by duality principle in projective geometry. Furthermore
Hilbert space  became a Procrustean bed even in quantum mechanics with finite degrees of freedom. Instead, one may mathematically describe the bra-ket  duality  in terms of  nuclearly rigged Hilbert spaces, aka  nuclear Gelfand triples  based on A. Grothendieck's  topological nuclear spaces (1955).  Applications of the nuclear L. Schwartz triples $\mathcal{S}(\mathbb{R}^n)\subset\mathcal{L}^2\subset\mathcal{S}^\prime(\mathbb{R}^n)$ to Gelfand-Kostyuchenko spectral expansion of self-adjoint  partial differential operators and to generalized random processes  have  been initiated by I. Gelfand already in 1955. From that time  the  systematic replacement of Hilbert spaces with nuclear triples (aka nuclearly rigged Hilbert spaces) became  a Gelfand doctrine.

The white noise analysis, launched by T. Hida  in 1975, acts on Gelfand triples
of Hida spaces of test and generalized functionals ( Dirac's  ket  and bra states) over Schwartz triples (see, e.g., \textsc{hida et al}\cite{Hida},  \textsc{obata}\cite{Obata-94}). 

\subsection{Polynomial operators}

An algebra  of finite-dimensional pseudodiferential operators,  introduced in \textsc{dynin}\cite{Dynin-61}, and refined by  J. Kohn and L. Nirenberg (1965) et al, has become a powerful tool for spectral theory of partial differential operators (see, e.g., \textsc{shubin}\cite{Shubin}).

Originally, the symbols were the normal ones but  J. Kohn and L. Nirenberg noticed  parallels with  Weyl symbols. In physics literature the latter were generalized by E. Wigner (1932) for quantum statistical mechanics, and then by  R. Glauber and E. Sudarshan (1963) for statistical quantum optics of new born lasers. In particular, E. Sudarshan introduced anti-normal symbols. A general theory of symbols (still in finite degrees of freedom and formal) was presented in 1970 by \textsc{agarwal-wollf}\cite{Agarwal}.

In quantum field a mathematical theory of  normal symbols was introduced in 1965 by \textsc{berezin}\cite{Berezin-65}.  By 1994 it was completed in the  T. Hida's white noise calculus framework of Gelfand nuclear triples (see \textsc{obata}\cite[Chapter 4]{Obata-94}) where polynomial operators are prominent.

In \textsc{berezin}\cite{Berezin-71}  anti-normal 
symbols were interpreted as compressed multiplication operators (still in finite dimensions) becoming a powerful tool in  theory of pseudodifferential operators.
A   generalization to quantum field theory was sketched in \textsc{dynin}\cite{Dynin-02}. In the present paper it  is completely refurbished.

\subsection{Quantization of non-linear systems}
This paper combines I. Segal's approach to  constructive  quantization of non-linear hyperbolic systems (see, e.g., (\textsc{segal}\cite{Segal-60})  with the quantization postulate of  \textsc{bogoliubov-shirkov}\cite[Section 9.4]{Bogoliubov}).

\smallskip
 The Segal's program  was to canonically  quantize the shell, i.e., solutions space of the partial differential system, rather than conventionally its classical   solutions. The basic idea  was that  the solutions space is a " differential manifold" with a natural symplectic form defined by the Cauchy data. \footnote{For starters,  he constructed  
Segal's prequantization of arbitrary simply connected \emph{finite-dimensional} symplectic manifold, the precursor of the powerful geometric quantization. }

It was suggested  the symplectic form may be the Peierls skew-symmetric  form of solutions for the tangential linear hyperbolic equations defined by their Green functions (see, e.g., \cite{DeWitt}).\footnote{ \textsc{segal}\cite{Segal-60}  
prequantized  arbitrary simply connected \emph{finite-dimensional} symplectic manifold, the start  of the powerful geometric quantization. }

Meanwhile \textsc{segal}\cite{Segal-79} established weak hyperbolicity of temporally gauged Yang-Mills equations, so that the solutions space may be parametrized by their \emph{constrained} initial data. As shown in the  present paper the constrained initial data form an \emph{infinite dimensional} K\"{a}ler manifold (as suggested by Segal himself).

\medskip
The classical  Yang-Mills energy-mass   functional  of the initial data is the time component of the   time-independent  Noether's energy-momentum functional. In the first order formalism the Gelfand triple (\ref{eq:CGelfand}) it becomes  the Yang-Mills Hamiltonian functional of canonically adjoint test fields  $(a,e^o)$. Note,   the usual constrains of the Hamiltonian (cp., e.g. \textsc{faddeev-slavnov}\cite[Section III.2]{Faddeev}) are resolved via the initial data trivialization. This leads to various ghostless 
quantum Yang-Mills Hamiltonians, the continuous linear operators on the test and generalized functionals on (\ref{eq:CGelfand}).  A conventional   normal quantization may produce operators that are not bounded from below even with non-negative normal symbols so that an infinite 
renormalization is necessary (see, e.g,  \textsc{glimm-jaffe}\cite{Glimm-69}).
 The infinite renormalization via  the anti-normal quantization achieves the semi-boundedness of the expectation functionals. 

Furthermore, the anti-normal quantum Hamitonian has  an infinite and discrete bosonic spectrum.

Conventional quantum field dynamics  looks for second quantized solutions of classical \emph{non-linear equations} with all ensuing problems of renormalization.
Feynman integration over classical histories without far reaching mathematical justification is just  a notation for divergent perturbation series. The  series terms
are computed via Feynman diagrams of bosonic particles interactions. Essentially,
this is a detour that is neither  a field theoretical, nor  quantum mechanical.

Instead,  following \textsc{schwinger}\cite {Schwinger-65} we consider   quantum dynamic of classical histories $z(t)\in\mathcal{C}_\mathbb{C}^\infty$ via the \emph{linear}  quantized Schr\"{o}dinger equation for the transition amplitudes
$\langle z(t)| z(0)\rangle\equiv \Omega_{ z(t)}^*\Omega_{ z(0)}$ 
\begin{equation}
\label{eq:Schr}
\frac{d}{dt}\langle z(t)| z(0)\rangle)\ =\ -iH\langle z(t)| z(0)\rangle.
\end{equation}
The equation has a unique solution in the form of a mathematically rigorous anti-normal Feynman type   integral over histories (see \textsc{dynin}\cite{Dynin-10}).
 
\subsection{Physics relations}
I. Segal's worked on  his  paper \cite{Segal-60} during his  stay at the University of Copenhagen (1959) in the footsteps of W. Heisenberg  announcement (1958) of his universal purely non-linear theory  of self-interaction of quantum spinor fields. Segal's goal was to overcome mathematical defects of Heisenberg 's work.

A similar approach (based on supersymmetric  Peierls-Poisson bracket) was undertaken by the physicist B. DeWitt whose lifework was has been summarized  in two volumes \textsc{dewitt}\cite{DeWitt}.  This monograph has rich contents but is not mathematically rigorous.

Segal himself never saw a mathematical completion of his program but was instrumental in igniting the  Cauchy  problem theory for classical  Yang-Mills fields
(see\textsc{segal}\cite{Segal-79}).

Recent physical papers by \textsc{frasca}\cite{Frasca} and \textsc{kholodenko}\cite{Kholodenko} gave intricate Higgsless arguments for existence of  a positive mass gap in QCD. Unfortunately, neither gave a complete argument for   the mathematical existence of quantum Yang-Mills theory.


\begin{thebibliography}{20}
\bibitem {Agarwal} 
 Agarwal, C. S., and Wolf, E. ,
\textit{Calculus for functions of noncommuting operators and general
phase-space methods in quantum mechanics}, Physical Rev. D \textbf{2}, 2161-2225, 1970. 

\bibitem{Berezin-65}
Berezin, F. A.,  \emph{The Method of Second Quantization}, Nauka. Moscow, 1965; Academic Press, 1966.

\bibitem{Berezin-71}
Berezin, F. A.,  \emph{Wick and anti-Wick operator symbols},
Mat. Sb., \textbf{86}, 578-610, 1971; Math. USSR  Sb., \textbf{15}, 577-606,  1971
 
 \bibitem{Berezin-91}
  Berezin, F. A., and Shubin, M. A.,   \emph{The Schr\"{o}dinger equation}, Kluwer Academic Publishers, 1991.
 
 \bibitem{Bogoliubov}
Bogoliubov, N. N., and Shirkov, D. V.,  \emph{Introduction to the Theory of Quantized Fields},  John Wiley, 1980.


\bibitem{Chen}
Chen, Y., et al, \emph{Glueball Spectrum and Matrix Elements on Anisotropic Lattices}, arXiv:hep-lat/0510074.
 

\bibitem{Dell'Antonio-91}
Dell'Antonio, G. and Zwanziger, D., \emph{Every gauge orbit passes inside the Gribov horizon},  Comm. Math. Phys., \textbf{138} (1991), 259-299. 

 \bibitem{DeWitt}  
DeWitt, B. S.,   \emph{The global approach to quantum field theory}, Clarendon Press, 2003.
  
 
\bibitem{Dirac-66}
Dirac, P., \emph{Lectures on quantum field theory}, Yeshiva University, New York,
1966.

\bibitem{Dynin-61}
Dynin, A.,  \emph{Singular integral operators of arbitrary order on manifolds},
Soviet Mthematics, Doklady, \textbf{2} (1961).

\bibitem{Dynin-02}
Dynin, A.,  \emph{Functional Weyl operators and Feynman Integrals}, Amer. Math. Soc. Translations ,  Series 2, \textbf{206},  (2002), p.65-80. arXiv:math-ph/0209057)

\bibitem{Dynin-10}
Dynin, A.,  \emph{Bosonization method for second super quantization}, J. of Nonlinear Mathematical Physics, \textbf{17}, Supplement 1 (2010), 1-13.
arXiv:0909.5160[math-phys]

\bibitem{Faddeev} 
 Faddeev,  L. D., and  Slavnov,A. A., \emph{Gauge Fields, Introduction to Quantum Theory},  Addison-Wesley, 1991.

\bibitem{Frasca}
Frasca, M.,   \emph{Glueball spectrum and hadronic processes in low-energy QCD} Nuclear Physics B - Proceedings Supplements, \textbf{207-208} (2010), 196-199.
(See also  M.Frasca's blog "The Gauge connection", December 01, 2010.)

\bibitem{Folland}
Folland, G. B, \emph{Harmonic Analysis in Phase Space},
Princeton University Press, 1989.

\bibitem{Gelfand}
 Gelfand, I. M., Vilenkin, N. Ya., \emph{Generalized Functions, Vol. 4, Applications to Harmonic Analysis}, Academic Press, 1964.
 
\bibitem{Glassey}
Glassey, R. T., and Strauss, W. A., \emph{Decay of Classical Yang-Mills fields}, Comm.. Math. Phys., \textbf{65} (1979), 1-13. 

\bibitem{Glimm-69}
Glimm, J.  and Jaffe, A., \emph{An infinite renormalization of the hamiltonian is necessary}, J. Math. Phys., \textbf{10} (1969), 2213-2214.

\bibitem{Goganov-85}
Goganov, M. V., and Kapitanskii, L. V., \emph{Global solvability of the Cachy problem for Yang-Mills-Higs equations}, Zapiski LOMI,
\textbf{147} (1985), 18-48; J. Sov. Math.,   \textbf{37} (1987), 802-822.


\bibitem{Hida}
Hida,T., Kuo, H.-H., Potthoff, J., Streit, L.,  \emph{White Noise: An Infinite Dimensional Calculus}, Kluwer, 1993 (Paperback 2010)


\bibitem{Clay-00}
Jaffe, A,  Witten, E., 
 http://www.claymath.org/millennium/Yang-Mills, 2000.


\bibitem{Kholodenko}Kholodenko, A., \emph{Gravity assisted solution of the mass gap problem for pure Yang-Mills fields}, \emph{International Journal of Geometric Methods in Modern Physics}, Accepted 2011-01-10 (arXiv:1001.0029).





\bibitem{Kuiper}
Kuiper, N.H., \emph{The homotopy type of the unitary group of a Hilbert space},Topology 
\textbf{3} (1965), 19-30.


\bibitem{Obata-94}
Obata, N.,  \emph{White Noise Calculus and Fock Space}, Lect. Notes
in Math., \textbf{1577}, Springer Verlag, 1994.


\bibitem{Reed}
Reed, M., Simon, B., \emph{Methods of Modern Mathematical Physics}, \textbf{II},
Academic Press, 1975.


\bibitem{Schwinger-65}
Schwinger, J., \emph{Nobel lecture "Relativistic quantum field theory"},  http://nobelprize.org, 1965.

\bibitem{Segal-60}
Segal, I., \emph{Quantization of nonlinear systems}, Journal of Mathematical Physics, \textbf{1}(1960), 468-488.

\bibitem{Segal-79}
Segal, I., \emph{The Cauchy Problem for the Yang-Mills Equations}, Journal of Functional Analysis, \textbf{33}(1979), 175-194.

\bibitem{Shubin} 
Shubin, M., \emph{Pseudodifferential operators and spectral theory}, Springer, 1987.


\bibitem{Strocchi}
 Strocchi, F., \emph{Selected Topics on the General Properties of 
Quantum Field Theory}, World Scientific, 1993.


\bibitem{Zhelobenko}
Zhelobenko, D. P., \emph{Compact Lie Groups and their Representations}, Providence, 1973.

 \end{thebibliography}
\end{document}